\documentstyle[sprocl,psfig]{article}

\def\linebreak{\unskip\break} 

\def\lcdm{$\Lambda$CDM}
\def\gsim{\mathrel{\raise.3ex\hbox{$>$\kern-.75em\lower1ex\hbox{$\sim$}}}}
\def\lsim{\mathrel{\raise.3ex\hbox{$<$\kern-.75em\lower1ex\hbox{$\sim$}}}}

\def\M10{{\times 10^{10} M_{\odot}\ }}

\def\hMpc{\ h^{-1}\ {\rm Mpc}}

\def\kms{\ {\rm km}\ {\rm s}^{-1}}
\def\kmsMpc{\ {\rm km}\ {\rm s}^{-1}\ {\rm Mpc}^{-1}}

\begin{document}

\title{Cosmological Structure Formation With and Without 
Hot Dark Matter\footnote{To appear in {\it Proceedings of the XVII
International Conference in Neutrino Physics and Astrophysics},
Neutrino 96,    Helsinki,
Finland 13-19 June 1996,
eds. K. Enqvist, K. Huitu and J. Maalampi (World Scientific, Singapore
1997).}}

\author{Joel R. Primack}
\address{University of California, Santa Cruz, CA 95064\\
{\it joel@physics.ucsc.edu}}

\maketitle

\abstracts{
The fact that the simplest modern cosmological theory, standard Cold
Dark Matter (sCDM), almost fits all available data has encouraged the
search for variants of CDM that can do better.  Cold + Hot Dark Matter
(CHDM) is the best theory of cosmic structure formation that I have
considered {\it if} the cosmological matter density is near critical
(i.e., $\Omega_0 \approx 1$) and {\it if} the expansion rate is not
too large (i.e. $h \equiv H_0/(100 \kmsMpc) \lsim 0.6$).  But I think
it will be helpful to discuss CHDM together with its chief competitor
among CDM variants, low-$\Omega_0$ CDM with a cosmological constant
(\lcdm).  While the predictions of COBE-normalized CHDM and \lcdm\
both agree reasonably well with the available data on scales of $\sim
10$ to $100 \hMpc$, each has potential virtues and defects. \lcdm\
with $\Omega_0 \sim 0.3$ has the possible virtue of allowing a higher
expansion rate $H_0$ for a given cosmic age $t_0$, but the defect of
predicting too much fluctuation power on small scales.  CHDM has less
power on small scales, and its predictions appear to be in good
agreement with data on the galaxy distribution, although it remains to
be seen whether it predicts early enough galaxy formation to be
compatible with the latest high-redshift data. Also, several sorts of
data suggest that neutrinos have nonzero mass, and the variant of CHDM
favored by this data --- in which the neutrino mass is shared between
two species of neutrinos --- also seems more compatible with the
large-scale structure data.  Except for the $H_0-t_0$ problem, there
is not a shred of evidence in favor of a nonzero cosmological
constant, only increasingly stringent upper bounds on it from several
sorts of measurements. Two recent observational results particularly
favor high cosmic density, and thus favor $\Omega=1$ models such as
CHDM over \lcdm\ --- (1) the positive deceleration parameter $q_0>0$
measured using high-redshift Type Ia supernovae, and (2) the low
primordial deuterium/hydrogen ratio measured in two different quasar
absorption spectra. If confirmed, (1) means that the cosmological
constant probably cannot be large enough to help significantly with
the $H_0-t_0$ problem; while (2) suggests that the baryonic
cosmological density is at the upper end of the range allowed by Big
Bang Nucleosynthesis, perhaps high enough to convert the ``cluster
baryon crisis'' for $\Omega=1$ models into a crisis for low-$\Omega_0$
models. I also briefly compare CHDM to other CDM variants such as
tilted CDM. CHDM has the advantage among $\Omega=1$ CDM-type models of
requiring little or no tilt, which appears to be an advantage in
fitting recent small-angle cosmic microwave background anisotropy
data.  The presence of a hot component that clusters less than cold
dark matter lowers the effective $\Omega_0$ that would be measured on
small scales, which appears to be in accord with observations, and it
may also avoid the discrepancy between the high central density of
dark matter halos from CDM simulations compared to evidence from
rotation curves of dwarf spiral galaxies.}

\section{Introduction}

``Standard'' $\Omega=1$ Cold Dark Matter (sCDM) with $h \approx 0.5$
and a near-Zel'dovich spectrum of primordial fluctuations~\cite{BFPR}
until a few years ago seemed to many theorists to be the most
attractive of all modern cosmological models. But although sCDM
normalized to COBE nicely fits the amplitude of the large-scale flows
of galaxies measured with galaxy peculiar velocity data
\cite{Dekel94}, it does not fit the data on smaller scales: it
predicts far too many clusters \cite{WEF93} and does not account for
their large-scale correlations \cite{Olivier93}, and the shape of the
power spectrum $P(k)$ is wrong \cite{BaughEf94,Zaroubi96}. Here I
discuss what are perhaps the two most popular variants of sCDM that
might agree with all the data: CHDM and \lcdm.  The linear {\it matter}
power spectra for these two models are compared in Figure~1 
with the real-space {\it galaxy} power spectrum obtained from the
two-dimensional APM galaxy power spectrum~\cite{BaughEf94}, which 
in view of the uncertainties is not in serious disagreement with
either model for $10^{-2} \lsim k \lsim 1 h$ Mpc$^{-1}$. The \lcdm\
and CHDM models essentially bracket the range of power spectra in
currently popular cosmological models that are variants of CDM.

\begin{figure}[htb]   
\vskip-1pc
\centering
\centerline{\psfig{file=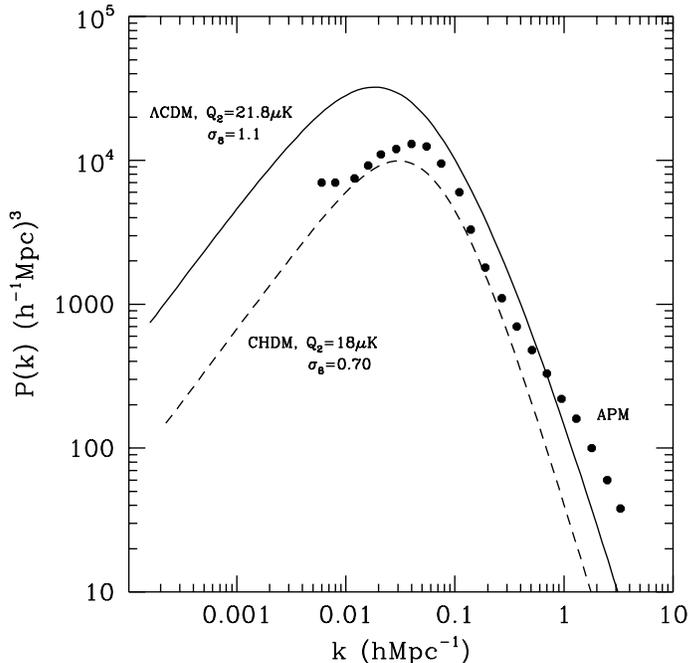,width=9cm}}
\vskip-1pc
\caption{%
Power spectrum of dark matter for \lcdm\ and CHDM models considered in
this paper, both normalized to COBE, compared to the APM galaxy
real-space power spectrum. (\lcdm: $\Omega_0=0.3$,
$\Omega_\Lambda=0.7$, $h=0.7$, thus $t_0=13.4$ Gy; CHDM: $\Omega=1$,
$\Omega_\nu=0.2$ in $N_\nu=2$ $\nu$ species, $h=0.5$, thus $t_0=13$
Gy; both models fit cluster abundance with no tilt, i.e. $n_p=1$.
From Ref.~\protect\cite{UCLA96}.)
}
\end{figure}

CHDM cosmological models have $\Omega=1$ mostly in cold dark matter
but with a small admixture of hot dark matter, light neutrinos
contributing $\Omega_\nu = m_{\nu,tot}/(92 h^2 {\rm eV}) \approx 0.2$,
corresponding to a total neutrino mass of $m_{\nu, tot} \approx 5$ eV
for $h=0.5$.  CHDM models are a good fit to much observational data
\cite{PHKC95,CHDM02} --- for example, correlations of galaxies and
clusters and direct measurements of the power spectrum $P(k)$,
velocities on small and large scales, and other statistics such as the
Void Probability Function (probability $P_0(r)$ of finding no bright
galaxy in a randomly placed sphere of radius $r$). My colleagues and I
had earlier shown that CHDM with $\Omega_\nu=0.3$ predicts a VPF
larger than observations indicate~\cite{Ghigna94}, but new results
based on our $\Omega_\nu=0.2$ simulations in which the neutrino mass
is shared equally between $N_\nu =2$ neutrino species~\cite{PHKC95} 
show that the VPF for this model is in excellent agreement with
observations \cite{Ghigna96}.  However, our simulations \cite{KPH96} of
COBE-normalized \lcdm\ with $h=0.7$ and $\Omega_0=0.3$ lead to a VPF
that is too large to be compatible with a straightforward
interpretation of the data \cite{Ghigna96}.  Acceptable \lcdm\  models
probably need to have $\Omega_0 >0.3$ and $h<0.7$, as discussed
further below.

Moreover, there is mounting astrophysical and laboratory data
suggesting that neutrinos have non-zero mass~\cite{PHKC95,FPQ}.  The
analysis of the LSND data through 1995~\cite{LSND96} strengthens the
earlier LSND signal for $\bar\nu_\mu \rightarrow \bar\nu_e$
oscillations. Comparison with exclusion plots from other experiments
implies a lower limit $\Delta m^2_{\mu e} \equiv
|m(\nu_\mu)^2-m(\nu_e)^2| \gsim 0.2$ eV$^2$, implying in turn a lower
limit $m_\nu \gsim 0.45$ eV, or $\Omega_\nu \gsim 0.02 (0.5/h)^2$.
This implies that the contribution of hot dark matter to the
cosmological density is larger than that of all the visible stars
($\Omega_\ast \approx 0.004$~\cite{Peebles}). More data and analysis
are needed from LSND's $\nu_\mu \rightarrow \nu_e$ channel before the
initial hint \cite{Cald95} that $\Delta m^2_{\mu e} \approx 6$ eV$^2$
can be confirmed.  Fortunately the KARMEN experiment has just added
shielding to decrease its background so that it can probe the same
region of $\Delta m^2_{\mu e}$ and mixing angle, with sensitivity as
great as LSND's within about two years.  The Kamiokande
data~\cite{KamAtmNu} showing that the deficit of $E > 1.3$ GeV
atmospheric muon neutrinos increases with zenith angle suggests that
$\nu_\mu \rightarrow \nu_\tau$ oscillations \cite{Whymutau} occur with
an oscillation length comparable to the height of the atmosphere,
implying that $\Delta m^2_{\tau \mu} \sim 10^{-2}$
eV$^2$~\cite{KamAtmNu} --- which in turn implies that if either
$\nu_\mu$ or $\nu_\tau$ have large enough mass ($\gsim 1$ eV) to be a
hot dark matter particle, then they must be nearly degenerate in mass,
i.e. the hot dark matter mass is shared between these two neutrino
species. The much larger Super-Kamiokande detector is now operating,
and we should know by about the end of 1996 whether the Kamiokande
atmospheric neutrino data that suggested $\nu_\mu \rightarrow
\nu_\tau$ oscillations will be confirmed and extended \cite{SKam}.
Starting in 1997 there will be a long-baseline neutrino oscillation
disappearance experiment to look for $\nu_\mu \rightarrow \nu_\tau$
with a beam of $\nu_\mu$ from the KEK accelerator directed at the
Super-Kamiokande detector, with more powerful Fermilab-Soudan,
KEK-Super-Kamiokande, and possibly CERN-Gran Sasso long-baseline
experiments later.

Evidence for non-zero neutrino mass evidently favors CHDM, but it also
disfavors low-$\Omega$ models.  Because free streaming of the
neutrinos damps small-scale fluctuations, even a little hot dark
matter causes reduced fluctuation power on small scales and requires
substantial cold dark matter to compensate; thus evidence for even 2
eV of neutrino mass favors large $\Omega$ and would be incompatible
with a cold dark matter density $\Omega_c$ as small as 0.3
\cite{PHKC95}. Allowing $\Omega_\nu$ and the tilt to vary, CHDM can
fit observations over a somewhat wider range of values of the Hubble
parameter $h$ than standard or tilted CDM~\cite{Liddle1}.  This is
especially true if the neutrino mass is shared between two or three
neutrino species \cite{PHKC95,Holtz89,HoltzPri93,PS95}, since then the
lower neutrino mass results in a larger free-streaming scale over
which the power is lowered compared to CDM; the result is that the
cluster abundance predicted with $\Omega_\nu \approx 0.2$ and $h
\approx 0.5$ and COBE normalization (corresponding to $\sigma_8 \approx
0.7$) is in reasonable agreement with observations without the need to
tilt the model\cite{Borgani96} and thereby reduce the small-scale
power further.  (In CHDM with a given $\Omega_\nu$ shared between
$N_\nu=2$ or 3 neutrino species, the linear power spectra are
identical on large and small scales to the $N_\nu=1$ case; the only
difference is on the cluster scale, where the power is reduced by
$\sim 20\%$ \cite{Holtz89,PHKC95,PS95}.)

Another consequence of the reduced power on small scales is that
structure formation is more recent in CHDM than in \lcdm. This may
conflict with observations of damped Lyman $\alpha$ systems in quasar
spectra, and other observations of protogalaxies at high redshift,
although the available evidence does not yet permit a clear decision
on this (see below). While the original $\Omega_\nu=0.3$ CHDM model
\cite{DSS92,KHPR93} certainly predicts far less neutral hydrogen in
damped Lyman $\alpha$ systems (identified as protogalaxies with
circular velocities $V_c \geq 50\kms$) than is observed \cite{DLAS,KBHP},
lowering the hot fraction to $\Omega_\nu \approx 0.2$ dramatically 
improves this~\cite{KBHP,Ma95}. Also, the evidence from
preliminary data of a fall-off of the amount of neutral hydrogen in
damped Lyman $\alpha$ systems for $z \gsim 3$~\cite{Storrie} is in
accord with predictions of CHDM~\cite{KBHP}.

However, as for all $\Omega=1$ models, $h \gsim 0.55$
implies $t_0 \lsim 12$ Gyr, which conflicts with age estimates from
globular cluster \cite{Chaboyer96} and white dwarf cooling
\cite{Wood}. The only way to accommodate both large $h$ and large
$t_0$ within the standard FRW framework of General Relativity is to
introduce a positive cosmological constant ($\Lambda>0)$ 
\cite{LLPR,CaPT}.  Low-$\Omega_0$ models with $\Lambda=0$ don't help
much with $t_0$, and anyway are disfavored by the latest small-angle
cosmic microwave anisotropy data \cite{CAT}.

\lcdm\  flat cosmological models with $\Omega_0 = 1 - \Omega_\Lambda 
\approx 0.3$, where $\Omega_\Lambda \equiv \Lambda/(3H_0^2)$, 
were discussed as an alternative to $\Omega=1$ CDM since the beginning
of CDM \cite{BFPR,Peeb84}.  They have been advocated more recently
\cite{LCDM} both because they can solve the $H_0-t_0$ problem and
because they predict a larger fraction of baryons in galaxy clusters
than $\Omega=1$ models. Early galaxy formation also is often considered to
be a desirable feature of these models. But early galaxy formation
implies that fluctuations on scales of a few Mpc spent more
time in the nonlinear regime, as compared with CHDM models. As has
been known for a long time, this results in excessive clustering on
small scales. My colleagues and I have found that a typical
$\Lambda$CDM model with $h=0.7$ and $\Omega_0=0.3$, normalized to COBE
on large scales (this fixes $\sigma_8\approx 1.1$ for this model), is
compatible with the number-density of galaxy clusters\cite{Borgani96},
but predicts a power spectrum of galaxy clustering in real space that
is much too high for wavenumbers $k=(0.4-1)h/{\rm Mpc}$~\cite{KPH96}.
This conclusion holds if we assume either that galaxies trace the dark
matter, or just that a region with higher density produces more
galaxies than a region with lower density. One can see immediately
from Figure~1 that there will be a problem with this \lcdm\ model,
since the APM power spectrum is approximately equal to the linear
power spectrum at wavenumber $k \approx 0.6 h$ Mpc$^{-1}$, so there is
no room for the extra power that nonlinear evolution certainly
produces on this scale (see Figure~1 of Ref.~\cite{KPH96} and further
discussion below).  The only way to reconcile the model with the
observed power spectrum is to assume that some mechanism causes strong
anti-biasing --- i.e., that regions with high dark matter density
produce fewer galaxies than regions with low density. While
theoretically possible, this seems very unlikely; biasing rather than
anti-biasing is expected, especially on small scales \cite{KNS96}.
Numerical hydro+N-body simulations that incorporate effects of UV
radiation, star formation, and supernovae explosions~\cite{YepesKKK}
do not show any antibias of luminous matter relative to the dark
matter.

Our motivation to investigate this particular \lcdm\  model was to
have $H_0$ as large as might possibly be allowed in the \lcdm\ class
of models, which in turn forces $\Omega_0$ to be rather small in order
to have $t_0 \gsim 13$ Gyr. There is little room to lower the
normalization of this \lcdm\  model by tilting the primordial power
spectrum $P_p(k)=A k^{n_p}$ (i.e., assuming $n_p$ significantly
smaller than the ``Zel'dovich'' value $n_p=1$), since then the fit to
data on intermediate scales will be unacceptable --- e.g., the number
density of clusters will be too small~\cite{KPH96}.  Tilted \lcdm\
models with higher $\Omega_0$, and therefore lower $H_0$ for $t_0
\gsim 13$ Gyr, appear to have a better hope of fitting the available
data, based on comparing quasi-linear calculations to the
data~\cite{KPH96,LiddleLCDM}. But all cosmological models with a
cosmological constant $\Lambda$ large enough to help significantly
with the $H_0-t_0$ problem are in trouble with new observations providing
strong upper limits on $\Lambda$ \cite{Primack96}: gravitational
lensing~\cite{Kochanek}, HST number counts of ellptical
galaxies~\cite{Driver}, and especially the preliminary results from
measurements using high-redshift Type Ia supernovae. The
analysis of the data from the first 7 of the Type Ia supernovae
from the LBL group \cite{Perl96b} gave $\Omega_0=1-\Omega_\Lambda=
0.94^{+0.34}_{-0.28}$, or equivalently $\Omega_\Lambda=
0.06^{+0.28}_{-0.34}$ ($<0.51$ at the 95\% confidence level).

It is instructive to compare the $\Omega_0=0.3$, $h=0.7$ \lcdm\ model
that we have been discussing with standard CDM and with CHDM.  At
$k=0.5 h$ Mpc$^{-1}$, Figs.~5 and 6 of Ref.~\cite{KNP96} show that the
$\Omega_\nu=0.3$ CHDM spectrum and that of a biased CDM model with the
same $\sigma_8=0.67$ are both in good agreement with the values
indicated for the power spectrum $P(k)$ by the APM and CfA data, while
the CDM spectrum with $\sigma_8=1$ is higher by about a factor of two.
CHDM with $\Omega_\nu=0.2$ in two neutrino species~\cite{PHKC95} also
gives nonlinear $P(k)$ consistent with the APM data (cf. Fig. 3 of
Ref. \cite{UCLA96}).

\section{Cluster Baryons}

I have recently reviewed the astrophysical data bearing on the values
of the fundamental cosmological parameters, especially $\Omega_0$
\cite{Primack96}.  One of the arguments against $\Omega=1$ that seemed
hardest to answer was the ``cluster baryon crisis''~\cite{WhiteFrenk}:
for the Coma cluster the baryon fraction within the Abell radius ($1.5
\hMpc$) is
\begin{equation}
f_b \equiv {M_b \over M_{tot}} \geq 0.009+0.050h^{-3/2},
\end{equation}
where the first term comes from the galaxies and the second from gas.
If clusters are a fair sample of both baryons and dark matter, as they
are expected to be based on simulations, then this is 2-3 times the
amount of baryonic mass expected on the basis of BBN in an $\Omega=1$,
$h\approx 0.5$ universe, though it is just what one would expect in a
universe with $\Omega_0 \approx 0.3$. The fair sample hypothesis
implies that
\begin{equation}
\Omega_0 = {\Omega_b \over f_b}
         = 0.33 \left({\Omega_b \over 0.05}\right)
           \left({0.15 \over f_b}\right).
\end{equation}

A review of the quantity of X-ray emitting gas in a sample of
clusters~\cite{Fabian} finds that the baryon mass fraction within
about 1 Mpc lies between 10 and 22\% (for $h=0.5$; the limits scale as
$h^{-3/2}$), and argues that it is unlikely that (a) the gas could be
clumped enough to lead to significant overestimates of the total gas
mass --- the main escape route considered in \cite{WhiteFrenk} (cf.
also~\cite{ClusBar}). If $\Omega=1$, the alternatives are then either
(b) that clusters have more mass than virial estimates based on the
cluster galaxy velocities or estimates based on hydrostatic
equilibrium \cite{BBlanchard} of the gas at the measured X-ray
temperature (which is surprising since they agree~\cite{BLubin}),
(c) that the usual BBN estimate $\Omega_b \approx 0.05 (0.5/h)^2$ is
wrong, or (d) that the fair sample hypothesis is wrong
\cite{MushotzkyL}.  Regarding (b), it is interesting that there are 
indications from weak lensing~\cite{Kaiser} that at least some
clusters may actually have extended halos of dark matter --- something
that is expected to a greater extent if the dark matter is a mixture
of cold and hot components, since the hot component clusters less than
the cold~\cite{BHNPK,KofmanKP}. If so, the number density of clusters
as a function of mass is higher than usually estimated, which has
interesting cosmological implications (e.g., $\sigma_8$ is a little
higher than usually estimated).  It is of course possible that the
solution is some combination of alternatives (a)-(d). If none of the
alternatives is right, then the only conclusion left is that $\Omega_0
\approx 0.33$.  The cluster baryon problem is clearly an issue that
deserves very careful examination.

It has recently been argued \cite{SSchramm} that CHDM models are
compatible with the X-ray data within observational uncertainties of
both the BBN predictions and X-ray data. Indeed, the rather high
baryon fraction $\Omega_b \approx 0.1 (0.5/h)^2$ implied by recent
measurements of low D/H in two high-redshift Lyman limit
systems~\cite{Tytler} helps resolve the cluster baryon crisis
for all $\Omega=1$ models --- it is escape route (c) above.  With the
higher $\Omega_b$ implied by the low D/H, there is now a ``baryon
cluster crisis'' for low-$\Omega_0$ models!  Even with a baryon
fraction at the high end of observations, $f_b \lsim 0.2
(h/0.5)^{-3/2}$, the fair sample hypothesis with this $\Omega_b$
implies $\Omega_0 \gsim 0.5 (h/0.5)^{-1/2}$.

\section{CHDM: Early Structure Troubles?}

Aside from the possibility mentioned at the outset that the Hubble
constant is too large and the universe too old for any $\Omega=1$
model to be viable, the main potential problem for CHDM appears to be
forming enough structure at high redshift. Although, as I mentioned
above, the prediction of CHDM that the amount of gas in damped Lyman
$\alpha$ systems is starting to decrease at high redshift $z \gsim 3$
seems to be in accord with the available data, the large velocity
spread of the associated metal-line systems {\it may} indicate that
these systems are more massive than CHDM would predict (see
e.g.,~\cite{Lu}).  Also, results from a recent CDM hydrodynamic
simulation \cite{HernDLAS} in which the amount of neutral hydrogen in
protogalaxies seemed consistent with that observed in damped Lyman
$\alpha$ systems led the authors to speculate that CHDM models would
produce less than enough; however, since the regions identified as
damped Lyman $\alpha$ systems in the simulations were not actually
resolved, this will need to be addressed by higher resolution
simulations for all the models considered.

Finally, Steidel et al.~\cite{Steidel} have found objects by their
emitted light at redshifts $z=3-3.5$ apparently with relatively high
velocity dispersions, which they tentatively identify as the
progenitors of giant elliptical galaxies. {\it Assuming} that the
indicated velocity dispersions are indeed gravitational velocities, Mo
\& Fukugita (MF)~\cite{MoF} have argued that the abundance of these
objects is higher than expected for the COBE-normalized $\Omega=1$
CDM-type models that can fit the low-redshift data, including CHDM,
but in accord with predictions of the \lcdm\ model considered
here. (In more detail, the MF analysis disfavors CHDM with $h=0.5$ and
$\Omega_\nu \gsim 0.2$ in a single species of neutrinos. They
apparently would argue that this model is then in difficulty since it
overproduces rich clusters --- and if that problem were solved with a
little tilt $n_p \approx 0.9$, the resulting decrease in fluctuation
power on small scales would not lead to formation of enough early
objects. However, if $\Omega_\nu \approx 0.2$ is shared between two
species of neutrinos, the resulting model appears to be at least
marginally consistent with both clusters and the Steidel objects even
with the assumptions of MF.  The \lcdm\ model with $h=0.7$ consistent
with the most restrictive MF assumptions has $\Omega_0 \gsim 0.5$,
hence $t_0 \lsim 12$ Gyr.  \lcdm\ models having tilt and lower $h$,
and therefore more consistent with the small-scale power constraint
discussed above, may also be in trouble with the MF analysis.) But in
addition to uncertainties about the actual velocity dispersion and
physical size of the Steidel et al. objects, the conclusions of the MF
analysis can also be significantly weakened if the gravitational
velocities of the observed baryons are systematically higher than the
gravitational velocities in the surrounding dark matter halos, as is
perhaps the case at low redshift for large spiral galaxies~\cite{NFW},
and even more so for elliptical galaxies which are largely
self-gravitating stellar systems in their central regions.

Given the irregular morphologies of the high-redshift objects seen in
the Hubble Deep Field \cite{vdB96} and other deep HST images, it seems
more likely that they are relatively low mass objects undergoing
starbursts, possibly triggered by mergers, rather than galactic
protospheroids.  Since the number density of the brightest of such
objects may be more a function of the probability and duration of such
starbursts rather than the nature of the underlying cosmological
model, it may be more useful to use the star formation or metal
injection rates \cite{Madau} indicated by the total observed
rest-frame ultraviolet light to constrain models \cite{SomPri}.  The
available data on the history of star formation 
\cite{Gallego,Lilly,Madau} suggests that most of the stars and most of
the metals observed formed relatively recently, after about redshift
$z\sim 1$; and that the total star formation rate at $z\sim 3$ is
perhaps a factor of 3 lower than at $z \sim 3$, with yet another
factor of $\sim 3$ falloff to $z \sim 4$ (although the rates at
$z\gsim 3$ could be higher if most of the star formation is in objects
too faint to see). This is in accord with indications from damped
Lyman $\alpha$ systems \cite{FallCP} and expectations for $\Omega=1$
models such as CHDM, but not with the expectations for low-$\Omega_0$
models which have less growth of fluctuations at recent epochs, and
therefore must form structure earlier.  But this must be investigated
using more detailed modelling, including gas cooling and feedback from
stars and supernovae \cite{SomPri}, before strong conclusions can be
drawn.

\section{Advantages of Mixed CHDM Over Pure CDM Models}

There are three basic reasons why a mixture of cold plus hot dark
matter works better than pure CDM without any hot particles: {\bf (1)}
the power spectrum shape $P(k)$ is a better fit to observations, {\bf
(2)} there are indications from observations for a more weakly
clustering component of dark matter, and {\bf (3)} a hot component may
help avoid the too-dense central dark matter density in pure CDM dark
matter halos. I will discuss each in turn.

{\bf (1) Spectrum shape.}  The pure CDM spectrum $P(k)$ does not fall 
fast enough on the large-$k$ side of its peak in order to fit
indications from galaxy and cluster correlations and power spectra.
This is also related to the overproduction of clusters in pure
CDM. The obvious way to prevent $\Omega=1$ sCDM normalized to COBE
from overproducing clusters is to tilt it a lot (the precise amount
depending on how much of the COBE fluctuations are attributed to
gravity waves, which can be increasingly important as the tilt is
increased).  But a constraint on CDM-type models that is likely to
follow both from the high-$z$ data just discussed and from the
preliminary indications on cosmic microwave anisotropies at and beyond
the first acoustic peak from the Saskatoon experiment
\cite{Netterfield} is that viable models cannot have much tilt, since
that would reduce too much both their small-scale power and the amount
of small-angle CMB anisotropy. As I have already explained, by
reducing the fluctuation power on cluster scales and below,
COBE-normalized CHDM naturally fits both the CMB data and the cluster
abundance without requiring much tilt.  The need for tilt is further
reduced if a high baryon fraction $\Omega_b
\gsim 0.1$ is assumed \cite{LiddleHib}, and this also boosts the 
predicted height of the first acoustic peak.  No tilt is necessary for
$\Omega_\nu=0.2$ shared between $N_\nu=2$ neutrino species with
$h=0.5$ and $\Omega_b=0.1$. Increasing the Hubble parameter in
COBE-normalized models increases the amount of small-scale power, so
that if we raise the Hubble parameter to $h=0.6$ keeping
$\Omega_\nu=0.2$ and $\Omega_b=0.1(0.5/h)^2=0.069$, then fitting the
cluster abundance in this $N_\nu=2$ model requires tilt $1-n_p \approx
0.1$ with no gravity waves (i.e., $T/S=0$; alternatively if
$T/S=7(1-n_p)$ is assumed, about half as much tilt is needed, but the
observational consequences are mostly very similar, with a little more
small-scale power). The fit to the small-angle CMB data is still good,
and the predicted $\Omega_{\rm gas}$ in damped Lyman $\alpha$ systems
is a little higher than for the $h=0.5$ case. The only obvious problem
with $h=0.6$ applies to any $\Omega=1$ model --- the universe is
rather young: $t_0=10.8$ Gyr.

{\bf (2) Need for a less-clustered component of dark matter.}  The
fact that group and cluster mass estimates on scales of $\sim 1
\hMpc$ typically give values for $\Omega$ around 0.1-0.2,
while larger-scale estimates give larger values around 0.3-1
\cite{Dekel94} suggests that there is a component of dark matter that
does not cluster on small scales as efficiently as cold dark matter is
expected to do.  In order to  quantify this, my colleagues and I have
performed the usual group $M/L$ measurement of $\Omega_0$ on small
scales in ``observed'' $\Omega=1$ simulations of both CDM and CHDM
\cite{NKP96}. We found that COBE-normalized $\Omega_\nu=0.3$ CHDM
gives $\Omega_{M/L} = 0.12-0.18$ compared to $\Omega_{M/L} = 0.15$ for
the CfA1 catalog analyzed exactly the same way, while for CDM
$\Omega_{M/L} = 0.34-0.37$, with the lower value corresponding to bias
$b=1.5$ and the higher value to $b=1$ (still below the COBE
normalization).  Thus local measurements of the density in $\Omega=1$
simulations can give low values, but it helps to have a hot component
to get values as low as observations indicate.  We found that there
are three reasons why this virial estimate of the mass in groups
misses so much of the matter in the simulations: (1) only the mass
within the mean harmonic radius $r_h$ is measured by the virial
estimate, but the dark matter halos of groups continue their roughly
isothermal falloff to at least $2r_h$, increasing the total mass by
about a factor of 3 in the CHDM simulations; (2) the velocities of the
galaxies are biased by about $70\%$ compared to the dark matter
particles, which means that the true mass is higher by about another
factor of 2; and (3) the groups typically lie along filaments and are
significantly elongated, so the spherical virial estimator misses
perhaps 30\% of the mass for this reason.  Our visualizations of these
simulations \cite{BHNPK} show clearly how extended the hot dark matter
halos are.  An analysis of clusters in CHDM found similar effects, and
suggested that observations of the velocity distributions of galaxies
around clusters might be able to discriminate between pure cold and
mixed cold + hot models \cite{KofmanKP}.  This is an area where more
work needs to be done --- but it will not be easy since it will
probably be necessary to include stellar and supernova feedback in
identifying galaxies in simulations, and to account properly for
foreground and background galaxies in observations.

{\bf (3) Preventing too dense centers of dark matter halos.} Flores
and I \cite{FP94} pointed out that dark matter density profiles with
$\rho(r) \propto r^{-1}$ near the origin from high-resolution
dissipationless CDM simulations \cite{CDMsims} are in serious conflict
with data on dwarf spiral galaxies (cf. also Ref. \cite{Moore}), and
in possible conflict with data on larger spirals \cite{FPBF93} and on
clusters (cf. \cite{Miralda,FP96}). Navarro, Frenk, \& White
\cite{NFW} agree that rotation curves of small spiral galaxies such as
DDO154 and DDO170 are strongly inconsistent with their universal dark
matter profile $\rho_{NFW}(r) \propto 1/[r(r+a)^2]$.  I am at present
working with Stephane Courteau, Sandra Faber, Ricardo Flores, and
others to see whether $\rho_{NFW}$ is consistent with data from high-
and low-surface-brightness galaxies with moderate to large circular
velocities are consistent with this universal profile. The failure of
simulations to form cores as observed in dwarf spiral galaxies either is a
clue to a property of dark matter that we don't understand, or is
telling us the simulations are inadequate. It is important to discover
whether this is a serious problem, and whether inclusion of hot dark
matter or of dissipation in the baryonic component of galaxies can
resolve it.  It is clear that including hot dark matter will decrease
the central density of dark matter halos, both because the lower
fluctuation power on small scales in such models will prevent the
early collapse that produces the highest dark matter densities, and
also because the hot particles cannot reach high densities because of
the phase space constraint \cite{TremaineGunn,KofmanKP}.  But this may
not be enough.

\section{Best Bet CDM-type Models}

As I said at the outset, I think CHDM is the best bet if $\Omega_0$
turns out to be near unity and the Hubble parameter is not too large,
while \lcdm\ is the best bet if the Hubble parameter is too large to
permit the universe to be older than its stars with $\Omega=1$.

Both theories do seem less ``natural'' than sCDM.  But although sCDM
won the beauty contest, it doesn't fit the data.  CHDM is just sCDM
with some light neutrinos.  After all, we know that neutrinos exist,
and there is experimental evidence --- admittedly not yet entirely
convincing --- that at least some of these neutrinos have mass,
possibly in the few-eV range necessary for CHDM.

Isn't it an unnatural coincidence to have three different sorts of
matter --- cold, hot, and baryonic --- with contributions to the
cosmological density that are within an order of magnitude of each
other?  Not necessarily.  All of these varieties of matter may have
acquired their mass from (super?)symmetry breaking associated with the
electroweak phase transition, and when we understand the nature of the
physics that determines the masses and charges that are just
adjustable parameters in the Standard Model of particle physics, we
may also understand why $\Omega_c$, $\Omega_\nu$, and $\Omega_b$ are so
close.  In any case, CHDM is certainly not uglier than \lcdm.

In the \lcdm\ class of models, the problem of too much power on small
scales that I discussed at some length for $\Omega_0=0.3$ and $h=0.7$
\lcdm\ implies either that there must be some physical mechanism that
produces strong, scale-dependent anti-biasing of the galaxies with
respect to the dark matter, or else that higher $\Omega_0$ and lower
$h$ are preferred, with a significant amount of tilt to get the
cluster abundance right and avoid too much small-scale power
\cite{KPH96}.  Higher $\Omega_0 \gsim 0.5$ also is more consistent with the
evidence summarized above against large $\Omega_\Lambda$ and in favor
of larger $\Omega_0$, especially in models such as \lcdm\ with
Gaussian primordial fluctuations.  But then $h\lsim0.63$ for $t_0
\gsim 13$ Gyr.

Among CHDM models, having $N_\nu=2$ species share the neutrino mass
gives a better fit to COBE, clusters, and small-scall data than
$N_\nu=1$, and moreover it appears to be favored by the available
experimental data \cite{PHKC95}.  But it remains to be seen whether
CHDM models can fit the data on structure formation at high redshifts.

\bigskip

\noindent{\bf Acknowledgments}  This work was partially supported
by NASA and NSF grants at UCSC. I thank my collaborators, especially
Anatoly Klypin, for many helpful discussions of the material presented
here.  Thanks also to Patrick Huet for comments on an earlier draft.

\def\mnras{MNRAS}
\def\aa{A\&A}
\def\apj{ApJ}
\def\apjs{ApJS}


\begin{thebibliography}{00}
\bibitem{Whymutau} The Kamiokande data is consistent with
atmospheric $\nu_\mu$ oscillating to any other neutrino species with a
large mixing angle. But (see discussion and references in, e.g.,
\cite{PHKC95,FPQ}) $\nu_\mu$ oscillating to $\nu_e$ with a large
mixing angle is probably inconsistent with reactor and other data, and
$\nu_\mu$ oscillating to a sterile neutrino $\nu_s$ (i.e., one that
does not interact via the usual weak interactions) with a large mixing
angle is inconsistent with the usual Big Bang Nucleosynthesis
constraints.
\end{thebibliography}
\end{document}